\documentclass{desyproc}
\newcommand {\pom} {I\!\! P}
\begin{document}
\title{Hard diffraction at CDF}

\author{{\slshape Konstantin Goulianos}\\[1ex]
The Rockefeller University, 1230 York Avenue, New York, NY 10065, USA\\[1ex]
(On behalf of the CDF Collaboration)}

\contribID{smith\_joe}


\acronym{EDS'09} 

\maketitle

\begin{abstract}
 We present a CDF measurement of diffractive dijet production in $\bar{p}p$ collisions at 1.96 TeV at the Fermilab Tevatron Collider  using data from an integrated luminosity of $\approx 310$~pb$^{-1}$ collected by triggering on a high transverse momentum jet in coincidence with a recoil antiproton detected in a roman pot spectrometer. We report final results for 4-momentum transfer squared $t>-4$~GeV$^2$, antiproton-momentum-loss fraction within 0.03-0.09, Bjorken-x of the interacting parton in the antiproton in the range 0.001-0.1, and jet transverse energies from 10 to 100 GeV. 
\end{abstract}

\section{Introduction}\label{sec:intro}

We present final results from a CDF measurement of single-diffractive (SD) dijet production in $\bar pp$ collisions at $\sqrt s=$~1.96~TeV at the Fermilab Tevatron Collider using data collected by triggering on a high transverse momentum jet in coincidence with a recoil antiproton detected in a Roman Pot Spectrometer (RPS)~\cite{Run2dijetPRD}. We consider proton diffractive dissociation, $\bar p+p\rightarrow \bar p+G_{\bar p}+X_p$, characterized by a rapidity gap (region of pseudorapidity~\footnote{Rapidity, $y=\frac{1}{2}\ln\frac{E+p_L}{E-p_L}$, and pseudorapidity, $\eta=-\ln\tan\frac{\theta}{2}$, where $\theta$ is the polar angle of a particle with respect to the proton beam ($+\hat z$ direction), are used interchangeably for particles detected in the calorimeters, since in the kinematic range of interest in this analysis they are approximately equal.} devoid of particles) adjacent to an escaping  $\bar p$, and  a final state $X_p$ representing  particles from the dissociation of the proton~\cite{Barone}. The rapidity gap, presumed to be caused by a color-singlet exchange with vacuum quantum numbers between the $\bar p$ and the dissociated proton, traditionally referred to as Pomeron ($\pom$) exchange, is related to $\xi_{\bar p}$, the forward momentum loss of the surviving $\bar p$, by $G_{\bar p}=-\ln\xi_{\bar p}$.

Several diffractive dijet results were obtained by CDF in Run~I~
\cite{jgj1995,jjRPS630}.
Among these, most striking is the observation of a breakdown of QCD factorization, expressed as a suppression by a factor of ${\cal{O}}(10)$ of the diffractive structure function (DSF) measured in dijet production relative to that derived from fits to parton  densities measured in diffractive deep inelastic scattering (DDIS) at the DESY $e$-$p$ collider HERA (see~\cite{jjRPS}). 
  
The present Run~II diffractive dijet measurement was performed in order to further characterize the diffractive structure function my measuring $t_{\bar p}$ distributions over a wide range of $t$ and jet transverse energy, $E_T^{\rm jet}$, namely $-t_{\bar p}\leq 4$~GeV$^2$ and $10^2<Q^2\approx (E_T^{\rm jet})^2<10^4$~GeV$^2$, and to search for diffractive dips. Below, we present the main results of this measurement and compare them with theoretical expectations.   
\section{Measurement}
These measurements were performed using the Run~II CDF detector and special data samples.  
\paragraph{Detector} 
\begin{wrapfigure}{r}{0.55\textwidth}
\vspace*{-1em}\includegraphics[width=0.55\textwidth]{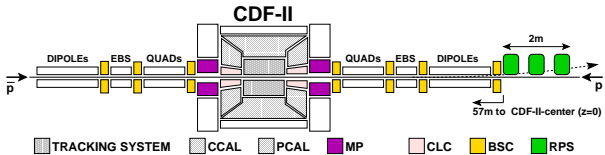}
\caption{\label{fig:fg1} 
Plan view of the CDF~II detector, showing the tracking system and calorimeters (central:CCAL, plug:PCAL, MiniPlugs:MP), the Cerenkov Luminosity Counters (CLC), and the Roman Pot Spectrometer (RPS): EBS are electrostatic beam separators.}
\vspace*{-2em}
\end{wrapfigure}

 Figure~\ref{fig:fg1} is a schematic plan view of the detector, showing the main CDF~II central detector and the forward detector-components essential to this measurement. The forward components include a Roman Pot Spectrometer (RPS), which measures $\xi_{\bar p}$ and $t_{\bar p}$ with resolutions $\delta\xi_{\bar p}=0.001$ and $\delta t_{\bar p}=\pm 0.07$~GeV$^2$ at $\left<-t_{\bar p}\right>\approx 0.05$~GeV$^2$, where $\delta t_{\bar p}$ increases with $t_{\bar p}$ with a $\propto\sqrt{-t_{\bar p}}$ dependence.
\paragraph{Data samples} This analysis is based on data corresponding to an integrated luminosity of $\cal{L}$$\approx 310$~pb$^{-1}$ collected in 2002--2003. Events were selected online with a three-level prescaled triggering system accepting RPS-triggered inclusive and jet-enriched events by requiring at least one calorimeter tower with $E_T>5,\,20,\mbox{ or }50$~GeV within  $|\eta|<3.5$. 
Jets were reconstructed using the midpoint algorithm~\cite{MidPoint}.

The majority of the data used in this analysis were recorded without RPS tracking information. For these data, 
the value of $\xi_{\bar p}$ was evaluated from calorimeter information and is designated as $\xi^{CAL}_{\bar p}$. The $\xi^{CAL}_{\bar p}$ was then calibrated against $\xi$ obtained from the RPS, $\xi^{RPS}_{\bar p}$,  using data from runs in which RPS tracking was available.

The following trigger definitions are used for these measurements:
\begin{itemize}
\renewcommand{\labelitemi}{$\bullet$}
\addtolength{\itemsep}{-0.5em}
\vspace*{-0.3em}
\item RPS: RPS trigger counters in time with a $\bar{p}$ crossing the nominal interaction point; 
\item J5 (J20, J50): jet with $E_T^{jet}\geq 5$ (20, 50) GeV in CCAL or PCAL;
\item RPS$\cdot$Jet5 (Jet20, Jet50): RPS trigger in coincidence with J5 (J20, J50).
\end{itemize}

\section{Results}
\begin{figure}[!htp]
\begin{center}
\includegraphics[width=0.44\textwidth]{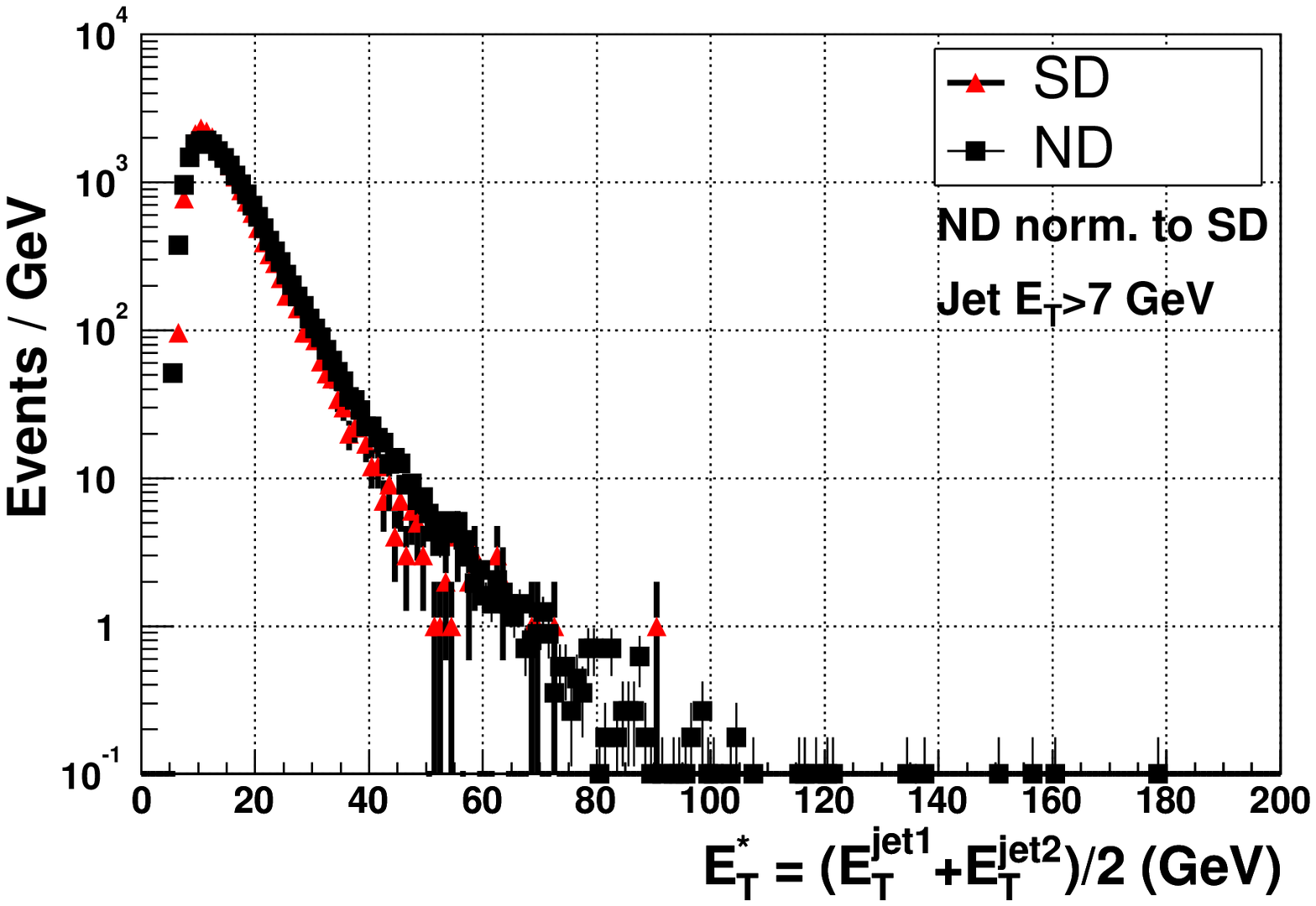}\hspace*{1em}\includegraphics[width=0.45\textwidth]{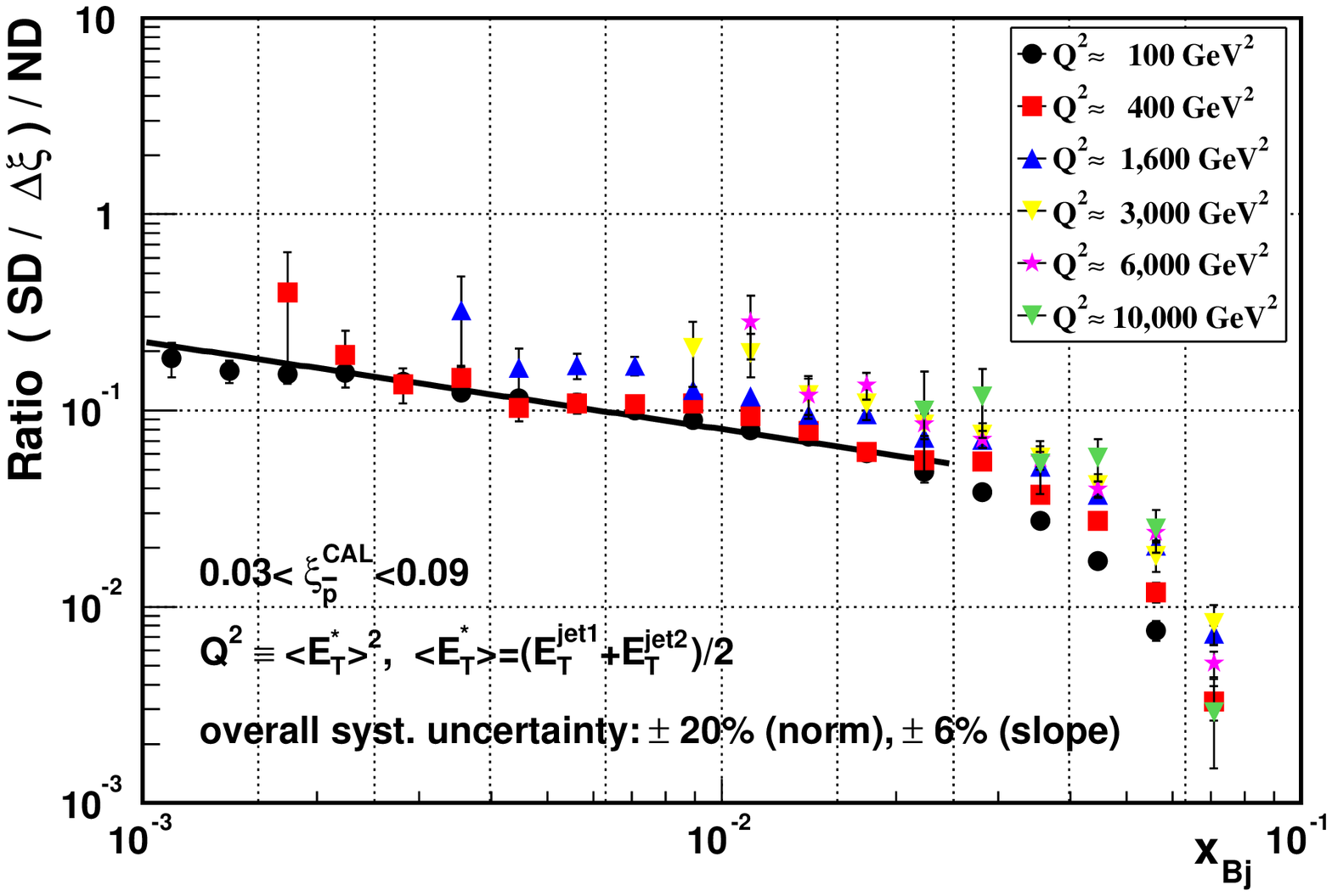}
\caption{\label{fig:fg2}
{\em (left)} Mean dijet transverse energy for SD and ND events normalized to the SD events; {\em (right)} ratios of SD to ND dijet-event rates vs $x_{\rm Bj}$ for various values of $\left<Q^2\right>\approx \left<E_T^*\right>^2$.}
\end{center}
\vspace*{-2em}
\end{figure}

In Fig.~\ref{fig:fg2}, we compare on {\em (left)} the mean dijet transverse energy between SD and ND events, 
and on {\em (right)} the $x_{\rm BJ}$ (Bjorken-$x$) distribution of the ratio of $({\rm SD}/{\Delta\xi})$/ND event-rates for various values of $\left<Q^2\right>\approx \left<E_T^*\right>^2$ over a range of two orders of magnitude. 
These plots show that the SD and ND distributions are very similar.
 
\begin{center}
\begin{figure}[!htb]
\includegraphics[width=0.35\textwidth]{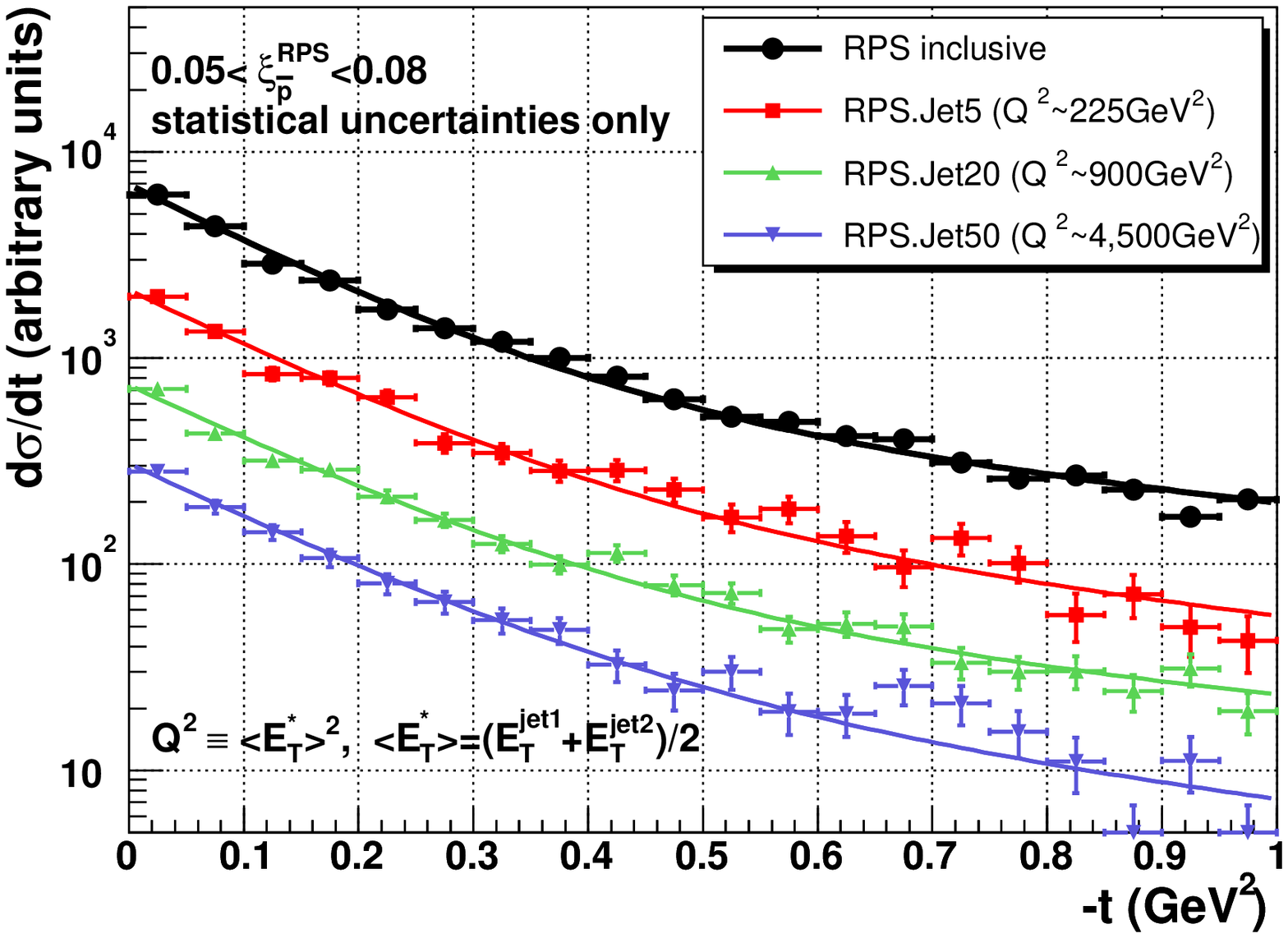}\hspace*{-1em}\includegraphics[width=0.35\textwidth]{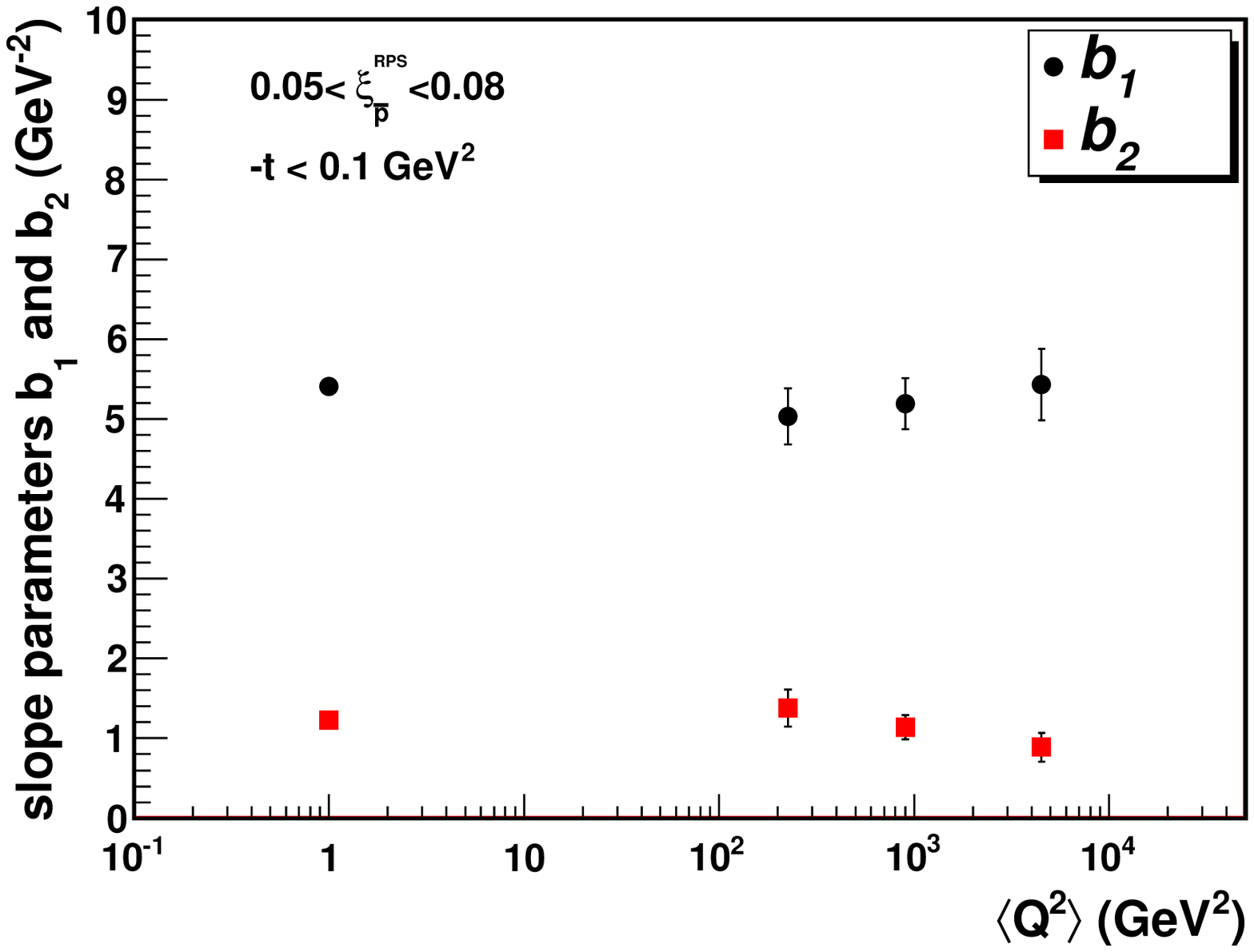}\hspace*{-1em}\includegraphics[width=0.35\textwidth]{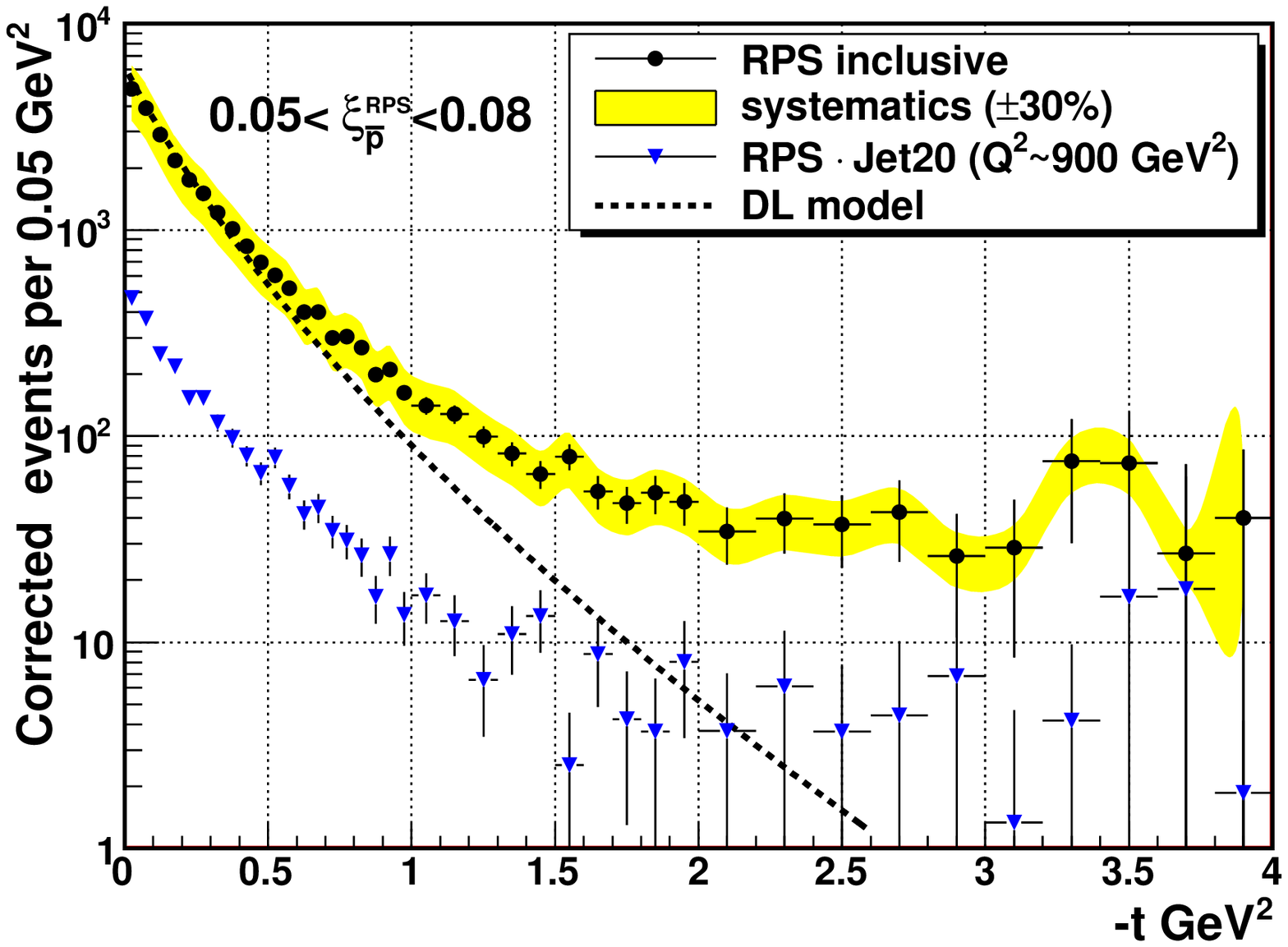}
\caption{\label{fig:fg3}
{\em (left)} $t_{\bar{p}}$ distributions of SD RPS data vs $\left<Q^2\right>$ for $0.05<\xi_{\bar{p}}^{\rm RPS}<0.08$; {\em (middle)} slope parameters $b_1$ and $b_2$ of a fit $dN_{\rm events}/dt=N_{\rm norm}(A_1e^{b_1t}+A_2e^{b_2t})$ with $A_2/A_1=0.11$ (see text) vs $\langle Q^2\rangle$; 
{\em (right)} $t$ distributions for RPS inclusive, $\left<Q^2\right>\simeq1\;{\rm GeV}^2$ (circles), and $\left<Q^2\right>\simeq900\;{\rm GeV}^2$ events (triangles) compared to the Donnachie-Landshoff (DL) model prediction.}
\end{figure}
\end{center}

The $t$ distributions for RPS inclusive and various dijet event samples are shown in Fig.~\ref{fig:fg3} (left) for $-t<1$~GeV$^2$ fitted to two exponential terms, and in Fig.~\ref{fig:fg3} (right) for $-t<4$~GeV$^2$. 
No significant variations are observed over a wide rage of $\langle Q^2\rangle$. For $-t<0.5$~GeV$^2$ all $t$ distributions, both for the inclusive and the high $\langle Q^2\rangle$ samples, are compatible with the expectation from the ``soft'' Donnachie-Landshoff (DL) model~\cite{ref:DL}. 
The rather flat $t$ distributions at large $-t$ shown in Fig.~\ref{fig:fg3} (right) are compatible with a possible existence of an underlying diffraction minimum around $-t\sim 2.5$~GeV$^2$ filled by $t$-resolution effects. These results favor models of hard diffractive production in which the hard
scattering is controlled by the parton-distribution-function
of the recoil antiproton while the rapidity-gap formation is
governed by a color-neutral soft exchange~\cite{Goulianos,Kopeliovitch}. 

\vspace*{-4em}
\section{Acknowledgments}
Warm thanks to my CDF colleagues and to the Office of Science of DOE for financial support.


\begin{footnotesize}

\end{footnotesize}

\begin{thebibliography}{99}
\bibitem{Run2dijetPRD}T. Aaltonen et al. (CDF Collaboration), Phys. Rev. D {\bf86}, 032009 (2012); arXiv:1206.3955 (2012).

\bibitem{Barone}V.~Barone and E.~Predazzi, ``High-Energy Particle Diffraction'', Springer Press, Berlin (2002).

\bibitem{jgj1995}F. Abe et al., (CDF Collaboration), 
Phys. Rev. Lett. {\bf 74}, 855 (1995).

\bibitem{jj}F. Abe et al. (CDF Collaboration), 
Phys. Rev. Lett. {\bf 79}, 2636 (1997).

\bibitem{jjRPS} T. Affolder et al. (CDF Collaboration), 
Phys. Rev. Lett. {\bf 84}, 5043 (2000).

\bibitem{jjRPS630} D. Acosta et al. (CDF Collaboration), 
Phys. Rev. Lett. {\bf 88}, 151802-(1-6) (2002).

\bibitem{MidPoint} G.~C.~Blazey et al., ``Run II Jet Physics'', in {\em Proceedings of the Run II QCD and Weak Boson Physics Workshop}; arXiv:hep-ex/0005012 (2000).
\bibitem{ref:DL}A.~Donnachie and P.~Landshoff, 
Phys. Lett. {\bf B518}, 63 (2001).

\bibitem{Goulianos}K.~Goulianos, ``Renormalized Diffractive Parton Densities,'' in {\em Diffraction 06, International Workshop on Diffraction in High-Energy Physics}, Adamantas, Milos island, Greece (2006), PoS (DIFF2006) 044 (2006). 



\bibitem{Kopeliovitch}B.Z. Kopeliovic {\em et al.,}
Phys. Rev. D {\bf 76}, 034019 (2007).
\end{thebibliography}
\end{document}